# STEM beam channeling in BaSnO₃/LaAlO₃ perovskite bilayers and visualization of 2D misfit dislocation network

Hwanhui Yun[1], Abhinav Prakash[1], Bharat Jalan[1], Jong Seok Jeong[1,2], K. Andre Mkhoyan[1*]

[1]*Department of Chemical Engineering and Materials Science, University of Minnesota, Minneapolis, MN 55455*

[2]*Analytical Sciences Center, LG Chem Ltd., Daejeon, Republic of Korea*

## Highlights

- STEM probe intensity depth profile is simulated to reveal the probe channeling in the BaSnO₃/LaAlO₃ bilayer.

- Defocus of a probe and atomic column arrangements in the bilayer modify channeling in a specimen.

- Depth of channeling is determined by specifics of examined materials unlike depth of field.

- Measured and simulated HAADF- and LAADF- STEM images of the BaSnO₃/LaAlO₃ bilayer visualize the MD network.

## Abstract

A study of the STEM probe channeling in a heterostructured crystalline specimen is presented here with a goal to guide appropriate STEM-based characterization for complex structures. STEM analysis of perovskite BaSnO₃/LaAlO₃ bilayers is performed and the dominating effects of beam channeling on HAADF- and LAADF-STEM are illustrated. To study the electron beam propagating through BaSnO₃/LaAlO₃ bilayers, probe intensity depth profiles are calculated, and the effects of probe defocus and atomic column alignment are discussed. Characteristics of the beam channeling are correlated to resulting ADF-STEM images, which is then tested by comparing focal series of plan-view HAADF-STEM images to those recorded experimentally. Additionally, discussions on how to visualize the misfit dislocation network at the BaSnO₃/LaAlO₃ interface using HAADF- and LAADF-STEM images are provided.

[*] **Corresponding author:** mkhoyan@umn.edu

**Introduction**

Scanning transmission electron microscope (STEM) is a powerful instrument for the analysis of structural, compositional, and electronic properties of crystalline materials at atomic scales. In STEM, a high-resolution projection image of atomic columns can be obtained from a zone-axis-oriented crystal by utilizing an annular dark-field (ADF) detector. In case of a single crystalline material, a STEM probe propagates through an atomic column that is uniform and comprised of one repeating unit throughout, and thus, an ADF-STEM image of the single crystal shows contrast that is representative of those repeating units. However, in some samples, structural variations along atomic columns are present, such as extended defects [1-8], voids [9-11], inclusions of different crystalline phase or composition [12, 13], misoriented grains [14], etc., and they need to be characterized. While depth sectioning [15-17] and quantification of ADF-STEM images [6, 10] have been proven to be successful methods for analysis of point defects, i.e. dopants and vacancies [1-11], there are not many reports on characterization of extended defects and layered crystals [17-19]. In STEM analysis of crystals, most of the characterization methods ultimately rely on understanding and utilization of probe channeling in the specimen [20-23], which in simple terms can be described as: the electron beam 'captured' by atomic potential propagating along an atomic column with specific oscillations defined by crystal structure of the sample. Therefore, the study of beam channeling in non-uniform crystalline specimens can guide the development of STEM-based characterization of complex structures.

Here, we study STEM probe channeling in the simplest of multi-section crystals – bilayer crystalline films – by evaluating and discussing its sensitivity on probe defocus and on misalignment of atomic columns in a sample. The resulting effects on ADF-STEM imaging of bilayers are outlined and implications toward visualization of the interface structures are discussed. The study is carried out on a bilayer of perovskite $BaSnO_3$ (BSO) and $LaAlO_3$ (LAO). This bilayer is an ideal model for such study, as it contains five different atomic columns and forms a variety of atomic column alignments between two layers. High-quality heteroepitaxial layers can be grown by vacuum deposition such as the molecular beam epitaxy method for STEM measurements [24]. Additionally, characterization of the heteroepitaxial interfaces of two perovskite oxides, including BSO-LAO [14], has been a topic of extensive research since discovery of two-dimensional (2D) electron gas at the interface between LAO and $SrTiO_3$ [25], and can benefit directly from such plan-view STEM study.

While rhombohedral LAO forms a pseudo-cubic perovskite structure along certain directions, BSO has a cubic perovskite structure [14]. The lattice mismatch between BSO and pseudo-cubic LAO is 8.0 %. Thus, when the heteroepitaxy of BSO and LAO is relaxed, misfit dislocations (MDs) are formed at the interface at every 11 u.c. of BSO (or 12 u.c. of LAO) [14]. The MD lines lie along the [100] and [010] crystallographic directions at the interface. Two sets of periodic MD lines cross each other and create two-dimensional (2D) MD network, as shown in Fig. 1. While these MDs have been observed in cross-sectional TEM and STEM images [14, 26-28], STEM-based plan-view characterization of these interfacial MD networks are lacking [18, 19]. This study provides an initial step in STEM characterization of these MD networks.

**Methods**

*Simulation*

The electron beam channeling through the bilayer is explored by calculating the intensity depth profiles of a STEM probe propagating through the BSO/LAO heteroepitaxial structure using the TEMSIM simulation package, which is based on the *Multislice* method [21, 23]. Supercell of the bilayer of BSO and LAO was constructed by stacking $11 \times 11 \times n_{BSO}$ unit cells of BSO on $12 \times 12 \times n_{LAO}$ pseudo-cubic unit cells of LAO ($n_{BSO}$ and $n_{LAO}$ are the number of BSO and LAO unit cells in perpendicular direction to the film surface, [001] direction.) There are two possible epitaxial alignments between these two layers: (1) aligning the AlO atomic column of LAO layer with the SnO column of BSO in the end unit cells, or (2) aligning the La atomic column of LAO with the Ba column of BSO in the end unit cells. As shown in Appendix, there is no substantial difference between these two models. All the simulations presented here are hence performed using the AlO-SnO alignment. With lattice parameters for pseudo-cubic LAO being 3.790 Å and for cubic BSO being 4.135 Å, supercells with size of $45.48 \times 45.48 \times (3.790 \times n_{LAO} + 4.135 \times$ $n_{BSO}$) Å$^3$ were constructed. This lattice constant of BSO allows boundary matching between LAO and BSO layers in the [100] and [010] directions.

STEM probe parameters were set as: beam energy of 200 keV, a convergence angle of 20 mrad, and defocus in the range from -8 nm to 32 nm. $C_s$ was set to be 0 as the differences between probes with small, aberration corrected $C_s$ and $C_s = 0$ are negligible [22, 29]. Thermal diffuse scattering was included using frozen-phonon approximation for T = 300 K [23]. Root-mean-square thermal displacements were set to be 0.0415 Å for La, 0.045 Å for Al, 0.0658 Å for O in LAO, 0.095 Å for Ba, 0.100 Å for Sn, and 0.089 Å for O in BSO [30]. ADF-STEM images were also simulated using the same TEMSIM package [21, 23]. The transmission functions were calculated using a 2048 × 2048 pixels$^2$ grid. High-angle ADF (HAADF)- and low-angle ADF (LAADF)-STEM image simulations were performed with detector inner/outer angles of 90/360 mrad and 20/80 mrad, respectively. The images were then convoluted with 2D Gaussian function with the full-width-at-half-maximum of 1 Å to incorporate the source size [31].

*Experiment*

BSO films were grown on pseudo-cubic LAO (001)$_{pc}$ substrates using hybrid molecular beam epitaxy [24]. Plan-view STEM samples were prepared by mechanical polishing (using Multiprep$^{TM}$ Allied High Tech Products, Inc.) followed by colloidal polishing with alumina abrasives to remove damaged layers on the surfaces. STEM experiments were carried out using aberration-corrected (probe-corrected) FEI Titan G2 60-300 (S)TEM operated at 200 keV. The probe convergent angle was 17.3 mrad and ADF detector inner angles were 93 and 19 mrad for HAADF- and LAADF-STEM images, respectively. The resolution was estimated using Au test specimen, and was around 0.8 Å. To remove high-frequency noise from the images, a low-pass filter was applied. The thicknesses of samples were determined from measured electron energy-loss spectra employing the log-ratio method [32] with the mean free paths of plasmon excitation: $\lambda_p$ = 81 nm for BSO and $\lambda_p$ = 89 nm for LAO [33].

**Results and discussion**

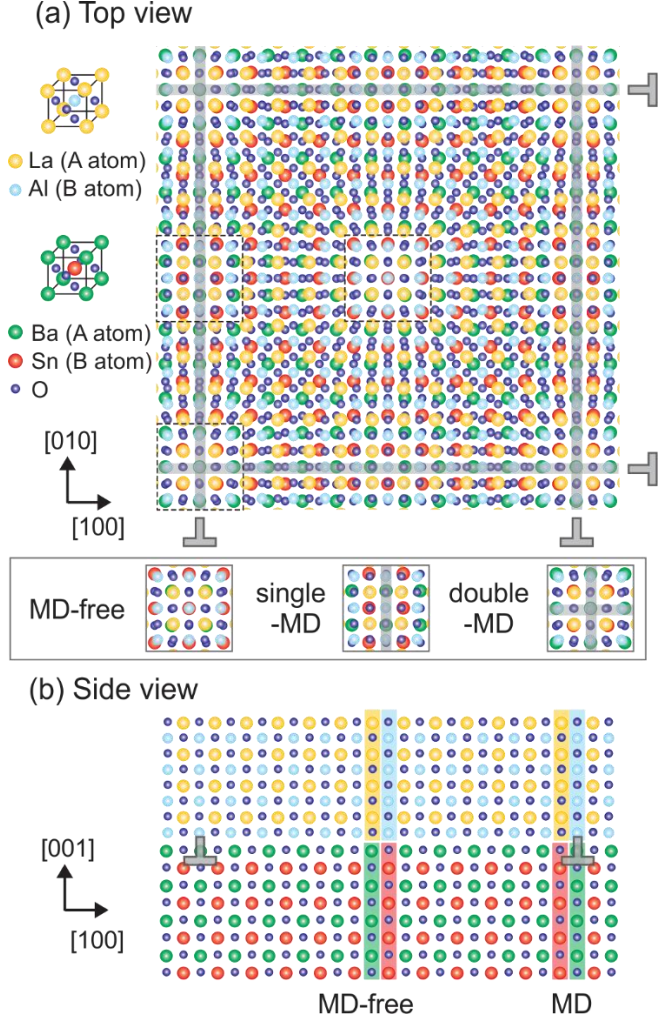

FIG 1. Schematic illustration of MD network for a bilayer of LAO on BSO. (a) MD network from top view (or plan-view). MD lines are marked with shaded lines. MD-free, single-MD, and double-MD regions are highlighted in a box below. (b) MD network from side view (or cross-sectional view). Position of MDs are marked with symbols in (a) and (b).

Schematics of the atomic arrangement in BSO/LAO bilayers from top view and the side view are shown in Figs. 1(a) and 1(b), respectively. Misalignment between atomic columns in LAO and BSO layers creates the Moiré-like pattern in the [001] projection with three distinct regions: MD-free, single-MD, and double-MD. While AlO-SnO and La-Ba alignments are formed in the MD-free region, the atomic columns in BSO and LAO are displaced from each other as they go away from the MD-free region. In the double-MD region, AlO-Ba and La-SnO alignments occur due to the half unit-cell lattice translation from one layer to another in the [010] and [100] directions. In single-MD regions, the half unit-cell deviation between two crystals only occurs in one direction, either in [010] or [100].

## 1. Channeling simulations

Since tracking the intensity of electron beam propagation through the film is essential for the understanding of beam channeling in bilayers, in this section, electron beam intensity depth profiles are calculated and analyzed for different STEM probes propagating through BSO/LAO bilayers.

### 1-1. Effects of probe defocus

Aberration-corrected STEM probe not only has a fine lateral size (< 1 Å), but also has dramatically reduced nano-scale focal depth [15, 20], thus, separating the effects of depth of field and defocus of the probe on beam channeling is necessary. While changing depth of field requires changes in probe convergence angle and other probe parameters, changing defocus is relatively straightforward and can be directly achieved during an operation of the STEM. To simplify discussion and directly compare results of calculations with experimental data, we first fixed probe parameters, i.e. the depth of field, and studied the effects of defocus.

Fig. 2 shows 2D electron intensity depth profiles of a probe propagating through a 30-nm-LAO/10-nm-BSO bilayer for different defocus values. Here, the probe was placed on the La-Ba column (Figs. 2(a) and (b)) and then on the AlO-SnO column (Figs. 2(c) and (d)) in the MD-free region. In Fig. 2(a), a focal series of the beam intensity depth profiles along the La-Ba column is shown. The average atomic number Z per distance in the La-Ba column is 15.04 Å$^{-1}$ and 13.54 Å$^{-1}$ for LAO and BSO layers, respectively, which puts beam channeling along these columns in high-Z regime [22]. In this regime, the electron beam propagates with strong intensity oscillations accompanied by fast decay. When the beam is focused at the sample entrance, i.e. $\Delta f = 0$, the beam channels through a La column in the top LAO layer and decays quickly. As the probe is focused away from the sample entrance, i.e. $\Delta f = 4, 8, 12$, and 16 nm, depth of channeling shifts down mimicking STEM depth sectioning [16, 17]. When the probe is defocused considerably away from the sample entrance, i.e. Δf > 12 nm, the not-fully-converged broader beam enters the sample and, as a result, a fraction of probe electrons is 'captured' by neighboring atomic columns.

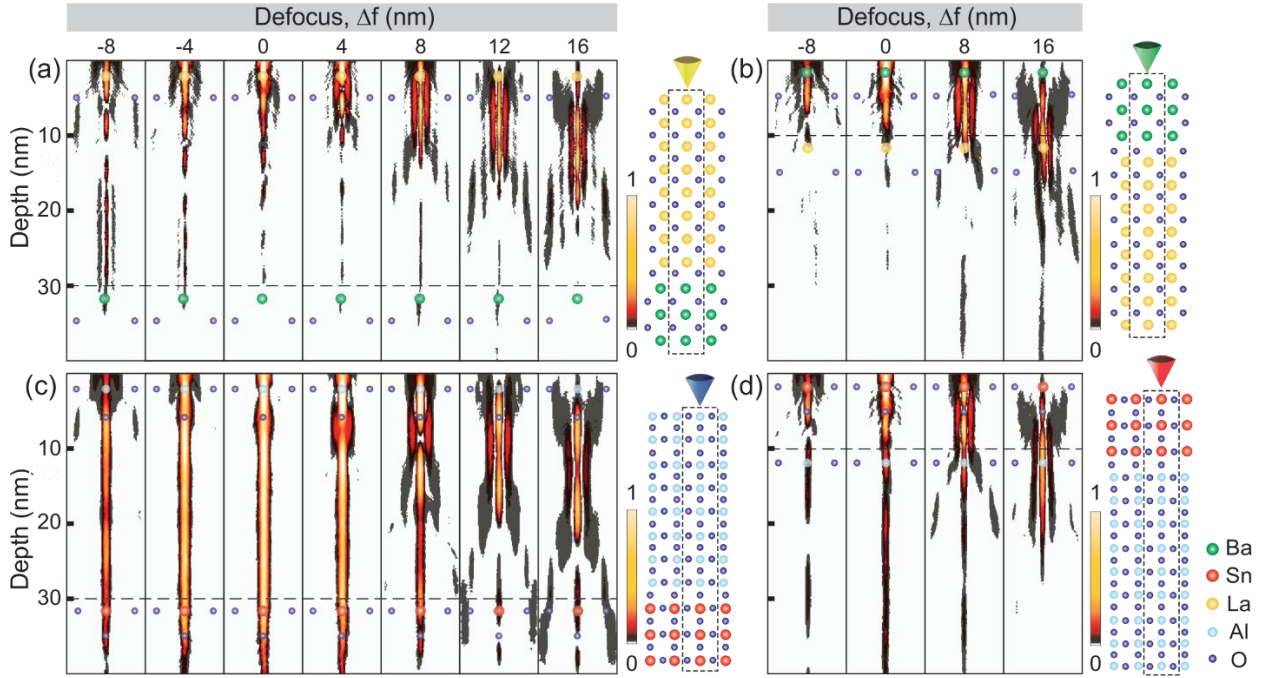

FIG 2. Focal series of 2D electron intensity depth profile for a STEM probe placed in the MD-free region of 30-nm-LAO/10-nm-BSO bilayer. (a, b) Focal series for a probe placed at the La-Ba column with beam direction: (a) LAO to BSO and (b) BSO to LAO. (c, d) Focal series for a probe placed at the AlO-SnO column with beam direction: (c) LAO to BSO and (d) BSO to LAO. Schematics of corresponding atomic planes are shown on the right of each depth profile. The atomic columns are indicated by circles in depth profiles. Each set of focal series was normalized using the maximum and minimum intensity of the series.

Even though the focal series resembles the depth sectioning, the depth of channeling and the depth of field of the probe are not the same. While the depth of the field is an intrinsic property of a probe, the depth of channeling is highly dependent on a crystal structure of a sample. Fig. 3 shows beam intensity depth profiles for a STEM probe placed in a vacuum and on the La column in LAO. As a probe channels along the La column, non-symmetric beam, which is comprised of several oscillation packets, is generated as shown in the extracted line profiles (bottom panel of of Fig. 3(b)). The depth of channeling at the La column is smaller than the depth of the field. The depth of channeling can be either smaller or considerably larger than depth of field depending on the linear density of an atomic column.

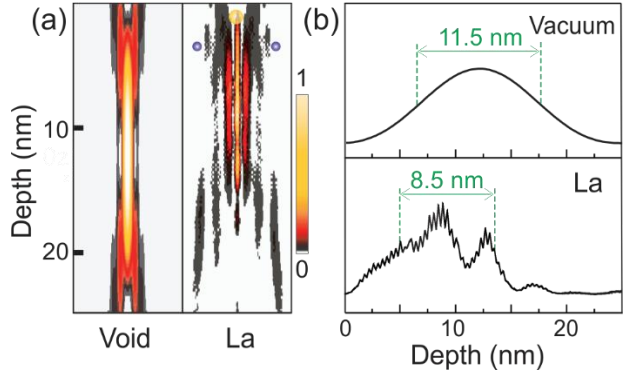

FIG 3. (a) 2D beam intensity depth profile of STEM probes propagating through a vacuum and a La column in LAO. Probe defocus was $\Delta f = 12$ nm. (b) Line profiles extracted from the optic axis of the depth profiles in (a). Depth of field (11.5 nm) and depth of channeling (8.5 nm) are estimated from the full-width-at-half-maximum of the profiles.

The intensity depth profile of the probe located on the AlO-SnO column shows very different defocus effect (Fig. 2(c)) than that in Fig. 2(a). The beam intensity decays slower than that along the La-Ba column due to the smaller average Z of the AlO column (5.54 Å$^{-1}$), which puts it into the low-Z regime [22]. This weak intensity decay along the AlO column extends the depth of channeling far beyond that of high-Z La-Ba columns, and even farther than the depth of field of the probe. At $\Delta f = -8$, -4, 0, 4, and 8 nm, the electron beam is no longer confined in the small region in a LAO layer, but channels across the interface. Interestingly, when the probe is defocused away from the entrance, $\Delta f = 12$, and 16 nm, the depth of channeling reduces and less electrons reach the Ba column in BSO layer.

When the sample is flipped, an incident beam enters to the thinner 10-nm-BSO layer first (Figs. 2(b) and 2(d)). Because Ba and SnO (average Z of 14.02 Å$^{-1}$) columns fall into high-Z regime, the depth of channeling is reduced for both columns. As defocus of the probe increases, the depth of channeling is modified, gets pushed into the middle of a specimen, and the distribution of the beam intensity shifts from BSO into the LAO. Interestingly, when the beam channels along the BaO-AlO column, once it reaches

LAO layer, the depth of channeling extends further into the layer channeling along the low-Z AlO column (Fig. 2(d)). These observations demonstrate that effects of the channeling cannot be underestimated when STEM depth sectioning is performed even on simpler bilayers.

In STEM, the electron beam intensity at a particular atomic column determines the amount of the electrons scattering from that column, generating signals for imaging and spectroscopy. HAADF-STEM image is formed by collecting the electrons incoherently scattered to higher angles, and the amount of the scattered electrons can be directly correlated to channeling beam intensity. Therefore, based on the simulated profiles, it can be speculated that when the beam enters from the 30-nm-LAO side, the HAADF-STEM images will be dominated by the LAO layer for a wide range of defocus values while extended depth of channeling at the AlO column will complicate quantification of these images. On the other hand, when the beam enters from the 10-nm-BSO side, contributions from each layer to HAADF-STEM image will depend on defocus of the probe.

*1-2. Effects of probe location*

As the electron beam propagates through two layers in the BSO/LAO bilayer, the channeling in a bottom layer is affected by beam propagation in the top layer. This is best illustrated by atomic column alignment between two layers. To examine this, the probe was placed at atomic columns in distinct regions of BSO/LAO bilayer: MD-free (①), single-MD (③,④), double-MD (⑥), in the middle of ① and ③ (②), and in the middle of ④ and ⑥ (⑤), as shown in Fig. 3(a). The beam propagation direction was set from BSO to LAO and defocus value was set $\Delta f = 8$ nm.

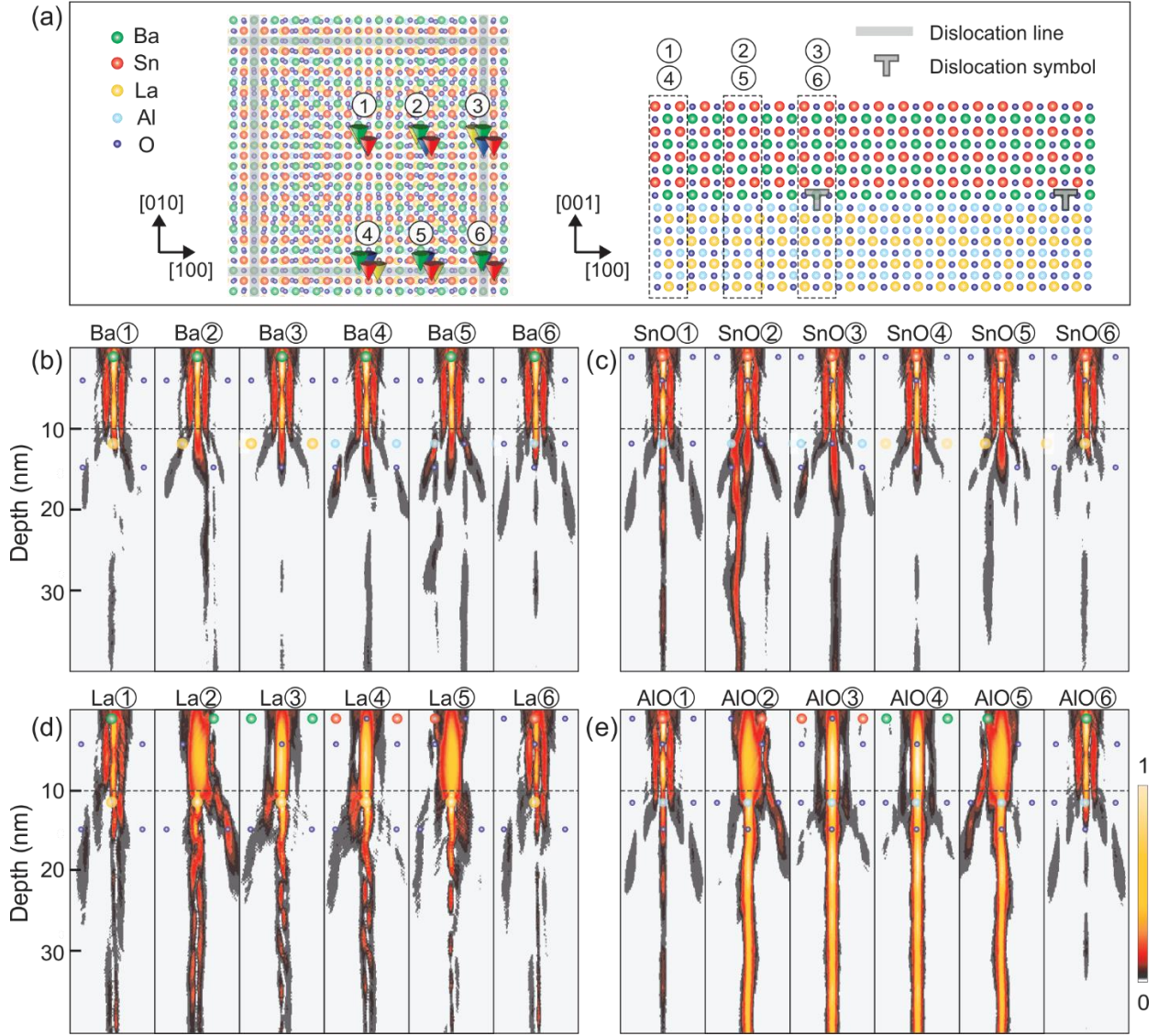

FIG 4. 2D electron intensity depth profile of a STEM probe propagating through the 10-nm-BSO/30-nm-LAO bilayer. (a) The probe positions for the depth profile simulations are displayed on atomic models from top-view and side-view. (b-e) Simulated depth profiles. Positions of atomic columns in the 2D plane is indicated with numbers in circles. The depth profiles were normalized using the same scale.

First, the probe was placed at Ba and SnO columns of the BSO layer (Figs. 4(b) and 4(c)). It should be noted that a beam channels along the columns in the first layer (in this case, BSO) the same way regardless of the region. In the MD-free region ①, since La and AlO atomic columns of the LAO layer are aligned with Ba and SnO columns, the beam continuously channels at the same position in the LAO layer, but with different decaying rate and oscillation wavelength. The beam channeling behavior in the double-MD region ⑥ is similar to that in the region ①. AlO and La columns are aligned with Ba and SnO columns, thus, the beam again channels at the same position but with different channeling intensity. In the single-MD regions (③ and ④), after channeling through the BSO layer, beam continues to channel along an oxygen column of the LAO layer located at the same position, but with much weaker channeling intensity.

In the regions ② and ⑤, atomic columns of BSO and LAO are misaligned. Thus, after the electron beam channels through Ba or SnO columns in the BSO layer, it strays away from the optic axis and transfers into neighboring atomic columns of the LAO layer. Such beam behavior in the second layer may cause unexpected results in ADF-STEM images as well as in spectroscopic elemental maps. Additionally, in cases of very small misalignment between atomic columns, through-column channeling in the first layer results in sub-atomic channeling in the second layer [29]. It is particularly evident in the depth profiles at the SnO column position in the region ② with clear sub-atomic channeling around the AlO column in the LAO layer (Fig. 4(c)).

Next, a STEM probe was located at the La and AlO column positions, on-column positions of the second layer (Figs. 4(d) and 4(e)). In the regions ① and ⑥, La and AlO columns are aligned with Ba and SnO columns of the first layer, and the results are the same as for the Ba and SnO positions in the regions ① and ⑥. In single-MD regions ③ and ④, the oxygen columns in the first layer weakly channel the electron through the BSO layer, which then channels strongly through the heavier La and AlO columns of the second LAO layer (Figs. 4(d) and 4(e)). Lastly, when a probe is placed at La and AlO columns in the regions ② and ⑤, the incident probe mostly propagates through an interatomic volume of the first layer with a small portion of the beam being captured by nearby atomic columns. As a result, the symmetry of the beam breaks and, as the beam enters to the LAO layer, sub-atomic channeling is invoked.

These electron intensity depth profiles demonstrate that the electron channeling in a bilayer (or in a multilayer) can be complicated since the electron channeling in a preceding layer strongly affects the channeling in successive layer. Also, a considerable electron beam intensity transfer can take place due to misaligned atomic columns.

## 2. HAADF-STEM image simulations

### 2-1. Defocus effect on HAADF-STEM images

Plan-view HAADF-STEM images were simulated for the BSO/LAO bilayers for a range of defocus values and film thicknesses with a goal to connect the electron beam intensity depth profiles to experimentally assessable images. In these image simulations, the thickness of a LAO layer was varied from 5 to 30 nm, while the thickness of a BSO layer was fixed at 10 nm. Fig. 5(a) shows the resulting HAADF-STEM images for the cases where the electron beam is propagating from BSO to LAO. As can be seen, when the beam enters a BSO layer first, the image contrast does not change with the LAO layer thickness, whereas the effects of defocus is significant. At $\Delta f = 0$ nm, the HAADF-STEM image shows contrast from both of BSO and LAO layers with the intensity from BSO being stronger than that from LAO. As defocus decreases and the probe is focused above the entry surface, the image becomes blurred due to probe spreading. On the other hand, as defocus increases, the depth of channeling moves toward the second layer and contribution from the LAO layer increases. At $\Delta f = 4$ and 8 nm HAADF-STEM images show clear contrast from both BSO and LAO layers with the characteristic Moiré-like pattern. At $\Delta f = 16$ nm, contrast from the BSO layer diminishes and mostly LAO is visible.

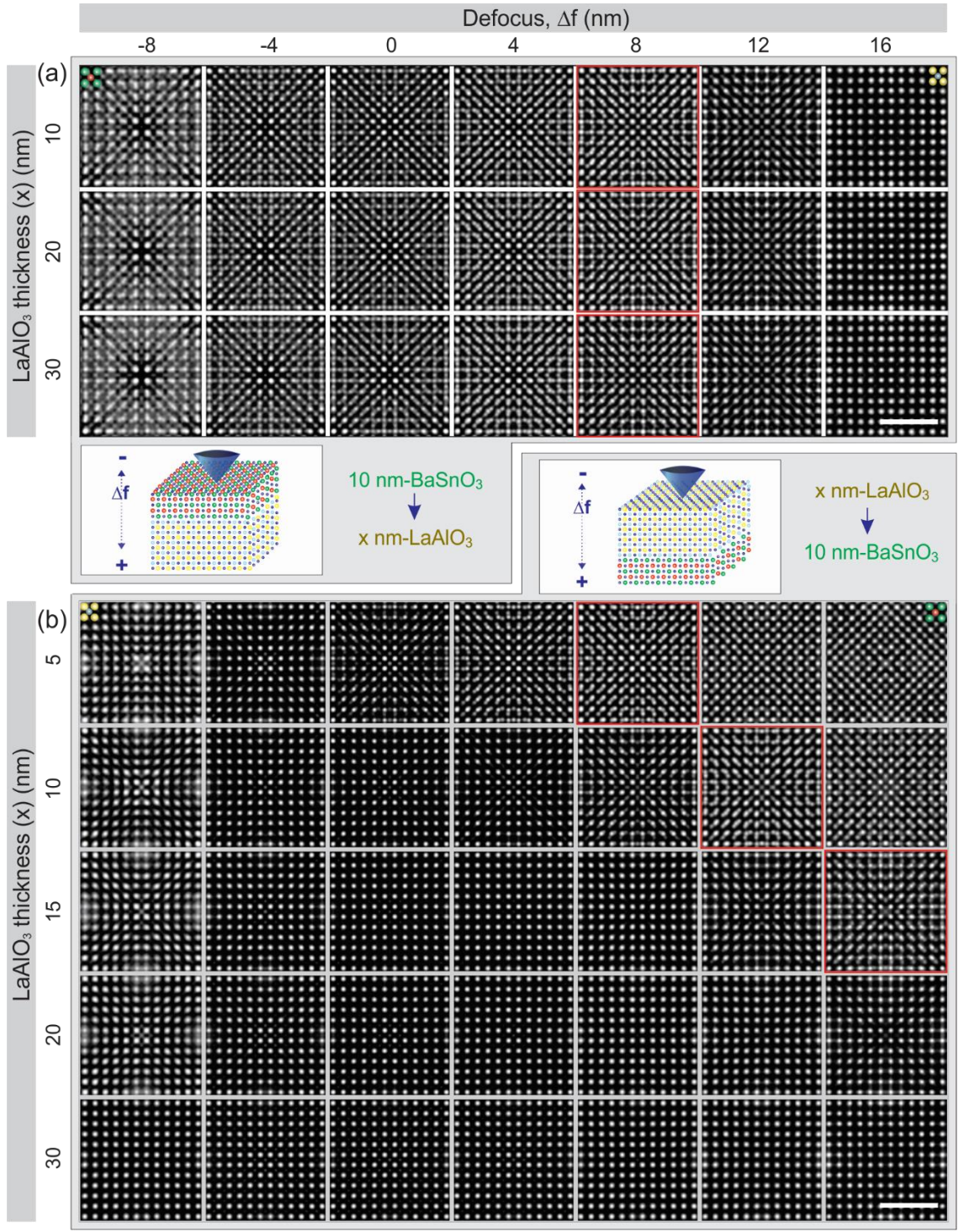

FIG 5. Simulated thickness-defocus map of plan-view HAADF-STEM images for BSO/LAO bilayers. The thickness of BSO was 10 nm and the thickness of LAO varied from 5 to 30 nm. The electron beam direction was (a) BSO to LAO and (b) LAO to BSO. A set of images displaying similar Moire-like pattern from the MD network are marked with a red box. Images were normalized individually, and scale bars are 2 nm.

HAADF-STEM images simulated for a beam propagating from LAO to BSO are shown in Fig. 5(b). The LAO thickness-defocus map shows strong dependence of image contrast on both the LAO layer thickness and defocus. For all LAO thicknesses, at $\Delta f$ = -8 to 0 nm HAADF-STEM images is dominated by contrast from the LAO layer. Here, slightly higher background observed at the single-MD regions in some of these LAO-dominant HAADF-STEM images is due to misaligned Ba and SnO columns in the bottom BSO layer. As defocus increases, contribution from the BSO layer adds up creating the characteristic Moiré-like pattern. Here, when the thickness of the LAO layer increases, the interface moves further away from the beam entry surface. Thus, increasing defocus shifts the depth of the channeling toward the deeper located interface and allows visualization of both BSO and LAO layers, as indicated by images in the red boxes (see Fig. 6 also). However, there are limitations to this. As defocus increases, a broader beam enters the sample causing weakening of main channeling and formation of satellite channelings in neighboring columns. Therefore, the interface depth that can be reached is restricted as too much defocus degrades the image. To illustrate this, Fig. 6(a) shows HAADF-STEM images of a 15-nm-LAO/10-nm-BSO bilayer simulated using $\Delta f$ = 0 and 16 nm and a 30-nm-LAO/10-nm-BSO bilayer using $\Delta f$ = 0 and 32 nm. The interface of the 15-nm-LAO and 10-nm-BSO can still be visualized using defocus $\Delta f$ = 16 nm. However, when the 30-nm-LAO/10-nm-BSO bilayer is imaged with $\Delta f$ = 32 nm, not only the contrast is dramatically lowered but the image resolution is degraded. Fig. 6(b) shows comparison of depth profiles of channeling probes with $\Delta f$ = 0 and 32 nm in the 30-nm-BSO/10-nm-BSO bilayer where drastic difference in beam intensity and degree of localization is visible.

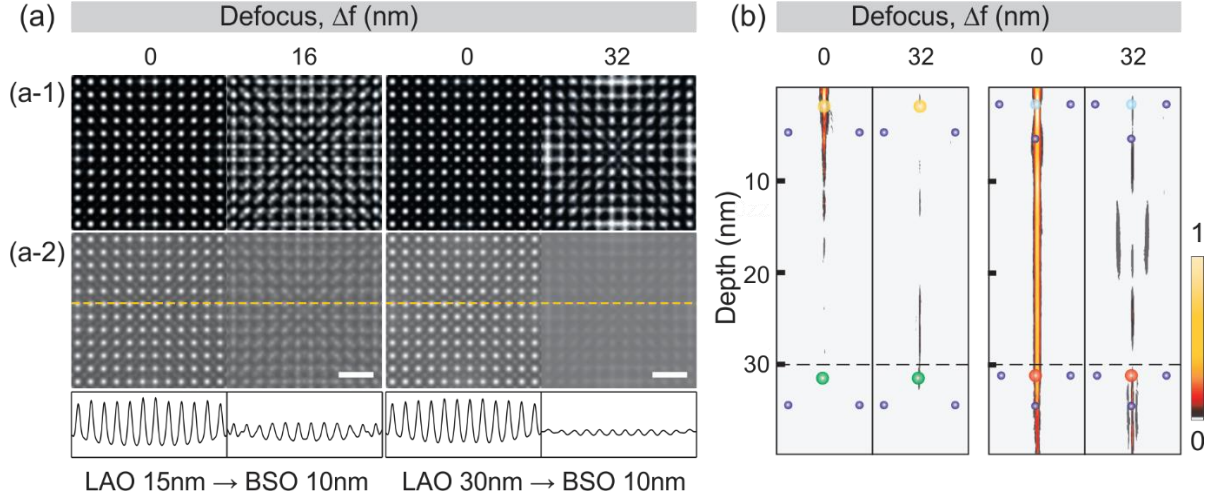

FIG 6. (a) HAADF-STEM images of the LAO/BSO bilayers with the electron beam propagation direction from LAO to BSO. (a-1) displays individually normalized images and (a-2) shows same images normalized to the incident probe intensity. Line profiles extracted from a dashed line in each image are presented below. Scale bars are 1 nm. (b) Beam intensity depth profile of a STEM probe propagating through 30-nm-LAO/10-nm-BSO bilayer. A probe is placed at La (left) and AlO (right) columns in the MD-free region and defocus values of 0 and 32 nm are used. Positions of the atomic columns are marked with closed circles.

## 2-2. *Comparison of simulations with experiments*

Focal series of HAADF-STEM images were experimentally acquired from a 10-nm-BSO/30-nm-LAO bilayer and compared with simulated images, as shown in Fig. 7. Two sets of experimental and simulated HAADF-STEM images were obtained with the beam propagating from BSO to LAO (Fig. 7(a)) and from LAO to BSO (Fig. 7(b)). As can be seen, measured and simulated images are in good agreement. As simulations predicted, in case of the beam propagating from BSO to LAO, the image contrast varies with probe defocus (Fig. 7(a)). When the defocus $\Delta f < 0$ nm, contrast from BSO is dominant. Then the LAO contrast is enhanced with increasing defocus. At $\Delta f = 4$ and 8 nm, 2D MD network is visible with the clear Moiré-like pattern. When the beam propagates from LAO to BSO, again, as simulations predicted, measured HAADF-STEM images are dominated by the LAO contrast regardless of defocus. Even slightly brighter contrast at the single-MD regions discussed earlier is observed in experiment. These results re-affirm the vital roles of probe defocus and the order in which layers are stacked on beam channeling and, consequently, on HAADF-STEM imaging of 2D interfacial MD networks.

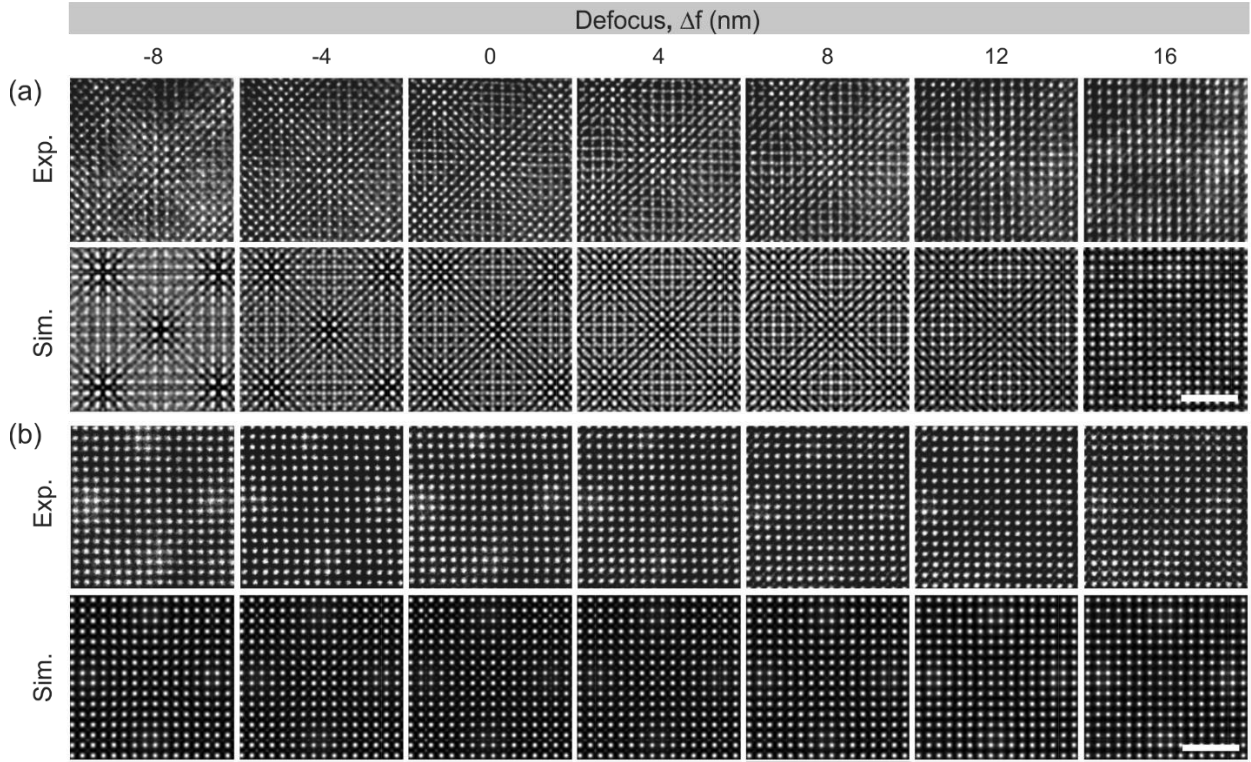

FIG 7. Measured and simulated focal series of plan-view HAADF-STEM images obtained from the 10-nm-BSO/30-nm-LAO bilayer. Images are for the incident electron beam propagating from (a) BSO to LAO and (b) LAO to BSO, respectively. Images were individually normalized. Scale bars are 2 nm.

*2-2. Imaging the MD network*

Unlike a HAADF-STEM image, a LAADF-STEM image of crystalline materials encompasses additional complexities due to overlapping low-order diffraction disks within LAADF detector angles. However, LAADF-STEM images can also provide useful information about crystal structure of a specimen [20, 34]. Pairs of experimental HAADF- and LAADF-STEM images were simultaneously obtained from the plan-view 10-nm-BSO/30-nm-LAO bilayer and are shown in Fig. 8 along with simulated images.

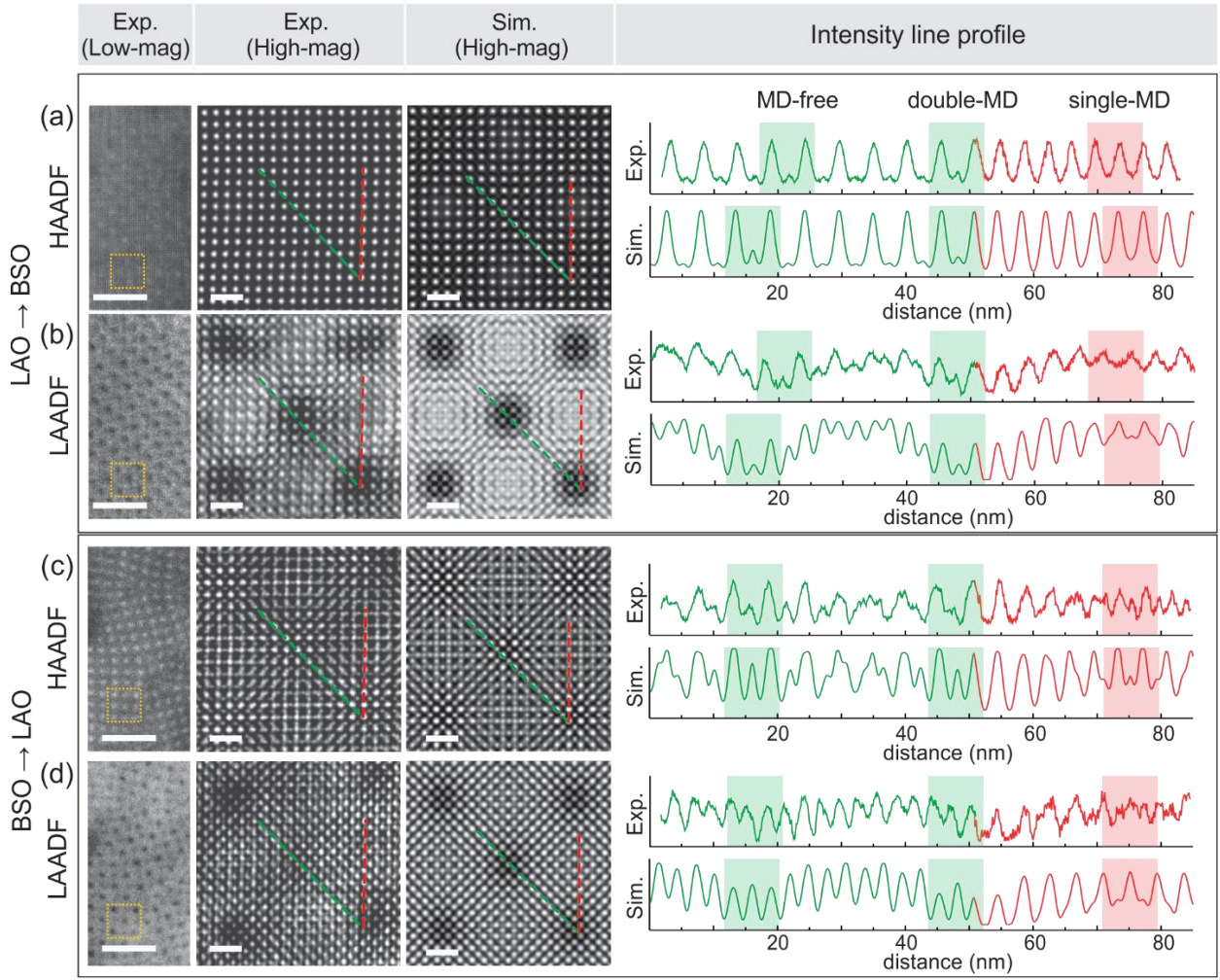

FIG 8. Measured and simulated pairs of HAADF-and LAADF-STEM images from 10-nm-BSO/30-nm-LAO bilayers. Beam propagation direction: (a,b) LAO to BSO, (c,d) BSO to LAO. The pairs of HAADF-and LAADF-STEM images were acquired simultaneously, and the images were individually normalized. Scale bars are 5 nm and 1 nm for low- and high-magnification images, respectively. The relative sizes of high-magnification images are indicated by yellow boxes overlaid on low-magnification images. The line profiles extracted from the high-resolution images are shown on the right. The profiles are from dashed lines shown in each image. MD-free and double-MD regions are indicated with green shades and single-MD region is with red shades.

When a beam propagates from LAO to BSO, a HAADF-STEM image shows weak bright contrast while a LAADF-STEM image exhibits strong dark contrast at low-magnification. High-magnification images (both measured and simulated) reveal that the bright contrast in a HAADF-STEM image is from the single-MD region, and the dark contrast in a LAADF-STEM image is from the MD-free and double-MD regions. Since the image simulations do not include strain in the structure, the strong dark contrast in the LAADF-STEM images is not from a strain field and should be explained otherwise. As a STEM probe propagates through a sample, the channeling modifies the angular distribution of the probe electrons by increasing the fraction of lower angle electrons [22]. As a result, LAADF-STEM images are sensitive to the electron scattering occurring deeper in the sample. Thus, in case of the 30-nm-LAO/10-nm-BSO bilayer, while a HAADF-STEM image is dominated by the top LAO layer, the LAADF-STEM image shows additional contrast from deeper located BSO layers. In particular, the enhanced sensitivity of the LAADF-STEM image on the bottom layer produces considerable background in the regions of the sample where atomic columns in bottom BSO are misaligned relative to those in LAO, making relatively darker contrast in the well-aligned atomic column regions. It should be noted that diffraction from the BSO layer also contributes to the LAADF-STEM image contrast and need to be considered for full quantification, but it is beyond the scope of this study.

Figs. 8 (c) and (d) show pairs of HAADF- and LAADF-STEM images from the 30-nm-LAO/10-nm-BSO bilayer with a probe propagating from BSO to LAO. Measured low-magnification HAADF image shows distinct contrast where bright spots are visible not only at the MD-free and double-MD regions but also at the single-MD regions. This observation is consistent with a prediction from simulations (see Fig. 5(b)). Here again, due to the enhanced sensitivity of a LAADF-STEM image on beam channeling, the regions with well-aligned atomic columns in the BSO/LAO bilayer exhibits relatively darker background intensity, which is distinctive even at low-magnifications. These observations show that LAADF-STEM images could be used to efficiently identify the presence of MD network at lower magnifications.

**Conclusion**

In this study, channeling of an aberration-free STEM probe through bilayers of heteroepitaxial perovskite BSO and LAO layers was explored. The effects of channeling on resulting HAADF- and LAADF-STEM images were evaluated, and then compared with experimental images. The effects of probe defocus and misalignment of atomic columns in two layers were analyzed. Resulting from this, discussion on limits of visualization of the MD network at the heteroepitaxial interface was provided. It was shown that, despite some similarities, depth of field and channeling width along atomic columns have substantial differences. While the former is purely defined by the STEM probe parameters, the latter has a strong dependence on specifics of crystalline structure of a sample. It was also shown that probe channeling in a multilayer is strongly affected by an atomic column alignment between subsequent layers. Therefore, interpretation of through focal series HAADF-STEM images of crystalline multilayers, such as crystals with extended defects and voids, different crystalline phase or misoriented grains, requires quantification of the beam channeling. Diverse electron channeling behaviors in sequential layers discussed in this study open new possibilities to manipulate beam propagation in a target material by using a layered sample. High sensitivity of LAADF signal on the beam channeling was also discussed, and it was shown that it can have some advantages over HAADF-STEM imaging by creating characteristic contrasts.

**Appendix**

There are two possible atomic column arrangements in the BSO/LAO bilayer. The schematics of both arrangements are illustrated in Fig. A1. In one case (model (A)), LAO and BSO layers are arranged to align the AlO column of LAO with the SnO column in BSO in the MD-free region. In other case (model (B)), LAO and BSO layers are arranged to align the La column of LAO with the Ba column in BSO in the MD-free region. As can be seen from simulated HAADF-STEM images the differences are negligible, which was expected considering very small differences in lattice constants of two crystals. With a BSO layer as a reference, the lattice of LAO layer in the model (A) is shifted in the [100] and [010] directions only by 0.1725 Å relative to the model (B).

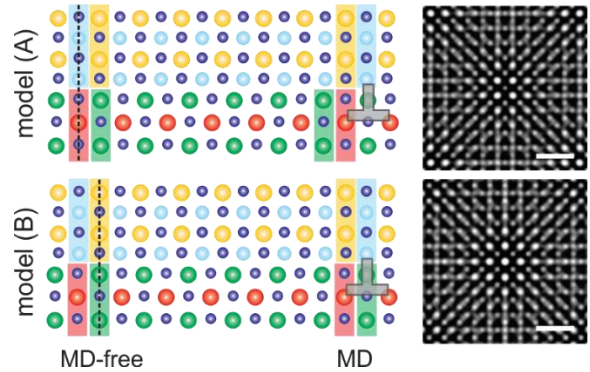

**Fig. A1.** Two possible arrangements of a bilayer of BSO and LAO. Atomic models from cross-sectional view are shown on the left and simulated plan-view HAADF-STEM images are shown on the right. Scale bars are 1 nm.

## Acknowledgement

This work is supported in part by SMART, one of seven centers of nCORE, a Semiconductor Research Corporation program, sponsored by National Institute of Standards and Technology (NIST), and by University of Minnesota (UMN) MRSEC program under award no. DMR-1420013. This work utilized the College of Science and Engineering (CSE) Characterization Facility, UMN, supported in part by the NSF through the UMN MRSEC program. Thin film growth work was supported by the National Science Foundation through DMR-1741801, UMN MRSEC program under award no. DMR-1420013 and through the Young Investigator Program of the Air Force Office of Scientific Research (AFOSR) through grant no. FA9550-16- 1-0205. H. Y. acknowledges a fellowship from the Samsung Scholarship Foundation, Republic of Korea.